\documentclass[twocolumn,showpacs,preprintnumbers,amsmath,amssymb]{revtex4-1}

\usepackage{graphicx}
\usepackage{dcolumn}
\usepackage{bm}

\def\3he{$^3$He}
\def\4he{$^4$He}

\begin{document}


\title{Mass Superflux in Solid Helium: the Role of $^3$He Impurities}

\author{Ye. Vekhov}
\author{R.B. Hallock}%
\affiliation{%
Laboratory for Low Temperature Physics, Department of Physics, University of Massachusetts, Amherst, Massachusetts 01003, USA
}%

\date{\today}

\begin{abstract}
Below $\sim 630$~mK, the \4he atom mass flux, $F$,
that passes through a cell filled with solid hcp \4he in the pressure range 25.6 - 26.4~bar, rises with falling temperature and at a temperature $T_d$ the flux drops sharply. The flux above $T_d$ has characteristics that are consistent with the presence of a bosonic Luttinger liquid. We study $F$ as a function of $^3$He concentration, $\chi = 0.17 - 220$~ppm, to explore the effect of $^3$He impurities on the mass flux. We find that the strong reduction of the flux is a sharp transition, typically complete within a few mK and a few hundred seconds. Modest concentration-dependent hysteresis is present.  We find that $T_d$ is an increasing function of $\chi$ and the $T_d(\chi)$ dependence differs somewhat from the predictions for bulk phase separation for $T_{ps}$ \emph{vs.} $\chi$. We conclude that $^3$He plays an important role in the flux extinction.  The dependence of $F$ on the solid helium density is also studied. We find that $F$ is sample-dependent, but that the temperature dependence of $F$ above $T_d$ is universal; data for all samples scales and collapses to a universal temperature dependence, independent of $^3$He concentration or sample history. The universal behavior extrapolates to zero flux in the general vicinity of $T_h \approx 630$~mK. With increases in temperature, it is possible that a thermally activated process contributes to the degradation of the flux. The possibility of the role of disorder and the resulting phase slips as quantum defects on one-dimensional conducting pathways is discussed.
\end{abstract}


\maketitle

\section{INTRODUCTION}
\label{introduction}

Solid helium is an unique substance that displays a combination of classical and quantum properties. It has been extensively studied both experimentally and theoretically for many decades.  One of the most interesting properties of \textit{liquid} helium is superfluidity, a state of matter in three dimensions than occurs below a pressure-dependent temperature $T_{\lambda}$. This quantum property is strongly affected by spatial limitation. In the two-dimensional (2D) case, the phase transition from the superfluid phase to the normal phase is related to the unbinding of vortices, as described by Berezinskii, Kosterlitz, and Thouless \cite{Berezinskii1972,Kosterlitz1973}. In one dimension (1D), another sort of quantum point defect, the so-called phase slip, is responsible for this transition. Quantum Monte-Carlo (QMC) simulation \cite{Boninsegni2007} predicted that the cores of screw dislocations in solid helium should be an example of 1D superfluidity. The flow of superfluid helium in 1D can be described by the quantum hydrodynamic theory known as Luttinger liquid theory \cite{Luttinger1963}. This idea has been confirmed  by large-scale QMC simulations \cite{DelMaestro2010,DelMaestro2011,Kulchytskyy2013} for the case of nanopores. The basic requirements for 1D channels to demonstrate Luttinger liquid behavior\cite{DelMaestro2012} are that the pore diameter is sufficiently small, the pore length is sufficiently long and the temperature is low enough with respect to $T_{\lambda}$.

We developed an apparatus, the so-called UMass Sandwich\cite{boris-06}, to study the possible ability of solid helium to carry a helium mass flux \cite{Ray2008a,Ray2009b,Vekhov2012}. Using porous media, Vycor rods, we are able to apply a chemical potential difference between two ends of a solid $^4$He sample without mechanically squeezing the solid helium lattice itself.  It was found that an experimental cell filled with solid \4he can carry a flux\cite{Ray2008a}, but only below some characteristic temperature, $T_h$, and the flux rate substantially increases with decreasing temperature.  Tiny amounts of the  impurity $^3$He also change the flux dramatically\cite{Vekhov2014b} at a characteristic low temperature, $T_d$. A brief report that discusses some of the \3he concentration dependence has appeared\cite{Vekhov2014b}. In this report we will describe our measurements as a function of the $^3$He concentration at several pressures in more detail and discuss our interpretations of the role of the $^3$He.

We note here that this report corrects a thermometry error that caused a small shift in the temperature scale below $\sim 100$ mK that was used in the work reported previously in Ref. \cite{Vekhov2014b}. This was caused by a change in the room temperature electronics which we subsequently determined introduced a small but measurable heating of the thermometer used to measure the solid helium temperature.  The temperature correction\cite{Vekhov2015x} for the work reported in Ref. \cite{Vekhov2014b} is less than 1 mk above 120 mK, 5.5 mK at 80 mK, and can be found from $TCnew = TC - 0.09637\exp(-TC/0.02755)$. $TC$ is the temperature of the thermometer affixed to the experimental cell, Fig.~\ref{cell_diagram_bw.eps}, and is used to define the solid helium temperature, $T$. All temperatures reported in this work include this correction.


\section{EXPERIMENTAL TECHNIQUE}
\label{technique}

In this work many freshly grown (and some partially annealed) solid \3he-\4he mixture samples have been used to study the effect of the $^3$He impurity concentration, $\chi$ (in the range $0.17 < \chi < 220$~ppm), temperature, and pressure on the \4he mass flux. Our experimental methods, have been described in substantial detail in Refs.\cite{Ray2008a,Ray2009b,Vekhov2012}. We provide a brief discussion of our approach here.

\subsection{Sample Preparation}

\begin{figure}[htb]
 \centerline{\includegraphics[width=0.9\linewidth,keepaspectratio]{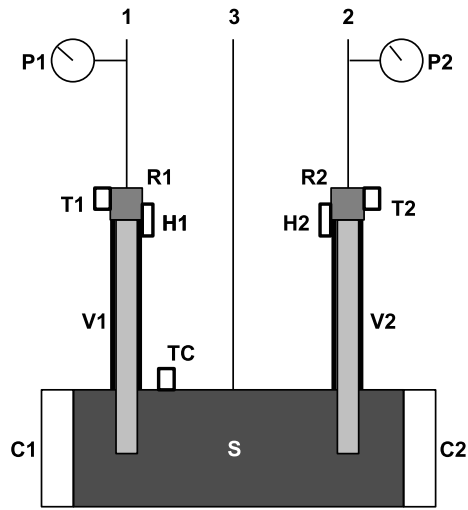}}%
\caption{ Schematic diagram of the cell used for
flow experiments.   Two capillaries, 1 and 2, go to
liquid reservoirs $R1$ and $R2$ at the top ends of the Vycor rods, $V1$ and $V2$.  Capillary 3 enters from the side of the cell and is used for adding helium to the cell.  Two capacitance
pressure gauges, $C1$ and $C2$, are located on either end of the cell for
\textit{in situ} pressure measurements of the solid \4he.  Pressures in the
lines 1 and 2 are read by pressure gauges, $P1$ and $P2$, outside the
cryostat.  Each reservoir has a heater, $H1$, $H2$, which prevents
the liquid in it from freezing and allows the temperatures of the reservoirs to be controlled.  The relevant temperatures are
read by calibrated carbon resistance thermometers $T1, T2$ and $TC$. [Reproduced from figure 1 in Ref. \cite{Ray2009b}] \label{cell_diagram_bw.eps}}
\end{figure}

Solid helium samples are grown at constant temperature from the superfluid in the  temperature range $0.3 < T < 0.4$~K by the condensation of helium into a sample cell (volume, $V = 1.84$~cm$^3$) through a direct-access heat-sunk capillary followed by an increase in the pressure up to near the melting pressure (about 25.34~bar). Subsequent additions of helium are by means of two other capillaries in series with Vycor (porous glass with
interconnected pores of diameter $\approx 7$~nm) rods ($0.14$~cm dia., $7.62$~cm long). Helium is added to the Vycor to inject atoms and create the solid at the desired pressure in the range of $25.6 < P < 26.4$~bar.  A cold plate at the base of the sample cell, Fig.~\ref{cell_diagram_bw.eps}, is thermally connected to the mixing chamber of dilution refrigerator and when filled with solid \4he can be cooled to about 60~mK. The lowest temperature of the cell is likely limited by the characteristics of our 1970's vintage SHE refrigerator and the heat flux through the superfluid-filled Vycor rods. Their warmer ends have to have much higher temperature than the temperature of the solid-filled cell, up to 1.5~K, to prevent the formation of solid helium in the two reservoirs R1 and R2 and at the interface between the Vycor rods and the reservoirs. The pressure range in the cell has an upper limit due to the need to maintain an adequate value of the superfluid density in the Vycor so as to not restrict the flux of superfluid helium through the Vycor.

To create samples of known \3he concentration, the cell is emptied between each set of measurements\cite{Vekhov2014b}. The cell is then filled with nominally pure \4he liquid (0.17 ppm \3he) up to the saturated vapor pressure through use of line 3.  Then, a small calibrated volume at room temperature is filled with pure \3he to a known pressure. This is injected into the cell via line 3 and this injection is followed by additional  \4he, which also enters through line 3, to bring the cell close to the melting curve. With knowledge of the relevant volumes and pressures, a known concentration of \3he is thus introduced into the cell.  The solid is then grown by injection of \4he through the two Vycor rods.  After the sample is grown, it is allowed to rest for $\approx 5-10$ hours at a solid helium temperature $\leq 0.4$~K before starting any measurements. Most solid helium samples are freshly grown (and not annealed above 0.5~K). As we will see, we find reproducibility in a given sample with temperature changes, which suggests that the samples are adequately in equilibrium after being created.  As has been seen previously for nominal-purity well helium (measured for this work to be $\sim$ 0.17 ppm \3he concentration), high temperature annealing leads, on cooling, either to a substantial flux decrease or to complete flux extinction, with in that case no evidence for flux at lower temperatures.

\subsection{Measurement Procedure}

To initiate the flux, an initial chemical potential difference, $\Delta \mu_0$, is applied between the tops of the Vycor rods by changing the temperatures, $T1$ and $T2$, of the two reservoirs, $R1$ and $R2$, to create a temperature difference between them (See Fig.~\ref{cell_diagram_bw.eps}). This creates a flux due to the fountain effect, which is seen by observing changes in P1 and P2.  Since we monitor the pressures and the temperatures as a function of time, the chemical potential $\Delta \mu$ can be calculated,

\begin{equation}
\Delta\mu = \int{\frac{dP}{\rho}}-\int{S dT},
\label{Eq_Dmu}
\end{equation}
\\
where $\rho$ and $S$ are the temperature-dependent density and entropy of liquid helium, respectively. In contrast to some of the earlier work from our lab, where $\Delta P = P1 - P2$ was applied \cite{Ray2009b} by direct mass injection, our current study uses the application of a temperature difference $\Delta T = T1 - T2$\cite{Ray2010b}. This approach offers two advantages. It allows for smaller density changes in the solid helium than was the case for direct injection of \4he to the sample cell through one of the lines, 1 or 2. And, it allows us to keep constant the total amount of \4he in the apparatus.

\begin{figure}[htb]
 \centerline{\includegraphics[width=1.1\linewidth,keepaspectratio]{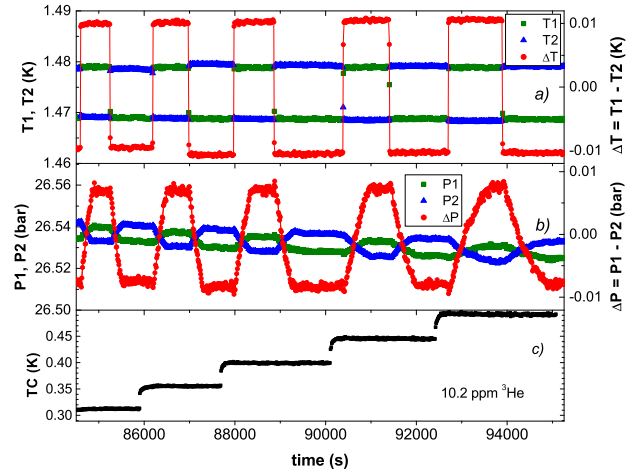}}%
\caption{(color online). An example of flux measurements for $10.2$~ppm sample. Here (a) temperatures are established for each of the reservoirs, held constant while the pressure in each of the reservoirs stabilizes, and then the temperature values of the two reservoirs are interchanged. The interchange results in a flux of atoms driven by the fountain effect which is recorded by (b) the pressure gauges $P1$ and $P2$.  The rate of change of $P1-P2$ provides a measure of the time dependent flux.  (c) The solid \4he temperature is changed and the process continues for a sequence of solid \4he temperatures, $TC$.  \label{Fig-10ppm-procedure}}
\end{figure}

An example of the procedure used for the flux measurements for a solid \4he sample with a $10.2$~ppm $^3$He impurity content is shown in Fig.~\ref{Fig-10ppm-procedure}. The creation of a change in the energy deposited in heaters $H1$ and $H2$ results in a temperature difference, $\Delta T = T1 - T2$, between the reservoirs Fig.~\ref{Fig-10ppm-procedure}(a), $R1$ and $R2$, at the tops of the Vycor rods and results in pressure responses Fig.~\ref{Fig-10ppm-procedure}(b), $P1$, $P2$ and $\Delta P = P1 - P2$ due to the fountain effect at a sequence of rising solid helium temperatures Fig.~\ref{Fig-10ppm-procedure}(c), $TC$. The derivative of $\Delta P$,

\begin{equation}
F = \frac{d(P1-P2)}{dt},
\label{Eq_F}
\end{equation}
\\
is taken to be reasonably proportional to the flux, $F$, of atoms that move from one reservoir to the other. We report the rate of pressure change in mbar/s units, where $0.1$~mbar/s corresponds to a flux of $\approx 4.8$ x $10^{-8}$~g/s.

We use measurements of $F$ of this sort in two related ways.  In one, we study how the flux, $F$, depends on the chemical potential, $\Delta\mu$, as time evolves, as the chemical potential changes from its initially imposed peak value, $\Delta\mu_0$, imposed by the initial $\Delta T$, to zero as equilibrium is restored by the creation of a fountain effect induced pressure difference, $P1-P2$.  In the other,  we study, for a given value of the imposed $\Delta T$, how the maximum resulting flux, $F$, depends on the temperature of the solid \4he.  Data, of the sort shown in Fig.~\ref{Fig-10ppm-procedure} for a specific 10.2 ppm \3he sample, is taken for a variety of solid helium samples, each with a specific value of the \3he concentration. We choose to focus on the behavior of $\Delta P$, which also allows us to eliminate a small long term drift in $P1$ and $P2$, which is typically present in our long-duration measurements due to main helium bath level changes. Our basic conclusions are not changed if instead we focus on the individual behaviors of $P1$ or $P2$.

\subsection{Flux Dependence on $\Delta \mu$}
\begin{figure}[htb]
 \centerline{\includegraphics[width=1.1\linewidth,keepaspectratio]{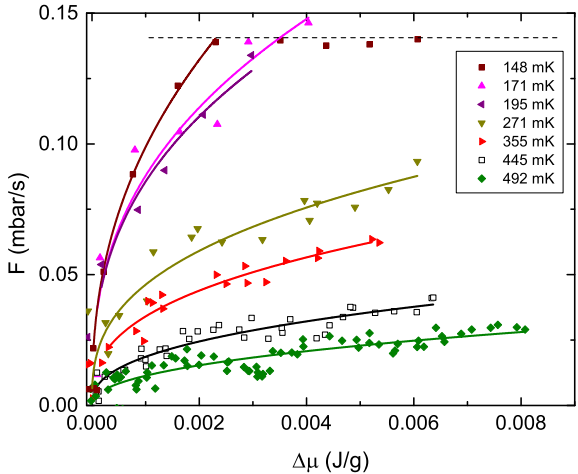}}%
\caption{(color online). An example of the $F(\Delta \mu)$ dependence for different solid helium temperatures for the case of a sample with a \3he concentration of 10.2 ppm. Solid lines are power law fits by use of Eq.(\ref{Eq-Fvsdmu}). The dashed line is the upper limit of the flux due to the Vycor bottle-neck for reservoir temperatures of $1.48$~K (see Fig.3 in Ref.\cite{Vekhov2014}). Note, before taking a derivative to calculate $F$, a moving average of the $\Delta P(t)$ data, Fig.\ref{Fig-10ppm-procedure}, was determined: by 3 points for $TC < 0.25$~K, by 7 points for $0.25 < TC < 0.40$~K, by 9 points for $TC = 0.445$~K and by 12 points for $TC = 0.492$~K. \label{Fig-10ppm-Fvsdmu}}
\end{figure}

Using data, including that in Fig.~\ref{Fig-10ppm-procedure}, $F$  \emph{vs.}  $\Delta \mu$ is obtained for positive $\Delta T$ values and presented in Fig.~\ref{Fig-10ppm-Fvsdmu} for a set of solid helium temperatures. The maximum flux values are typically constrained by the solid helium sample. But, for the lower temperatures, the constraint is imposed by the temperature of the reservoir at the upper end of the Vycor as shown previously\cite{Vekhov2014}. The dashed line in Fig.~\ref{Fig-10ppm-Fvsdmu} represents the flux limit imposed by the Vycor.  Similar behavior is seen for negative $\Delta T$ values.  As was found earlier\cite{Vekhov2012}, a power law provides   a good two-parameter characterization for data of this sort:

\begin{equation}
F = A (\Delta \mu)^b,
\label{Eq-Fvsdmu}
\end{equation}
\\
where $A$ and $b$ are fit parameters. The parameter $b$ is temperature independent and as we will see, in the pressure range of our study $b$ is less than 0.5 \cite{Vekhov2012}, but does depend on pressure. We will return to a discussion of the characteristics of $A$ and $b$ later.

\subsection{Flux Dependence on Temperature}

\begin{figure}[htb]
 \centerline{\includegraphics[width=1.1\linewidth,keepaspectratio]{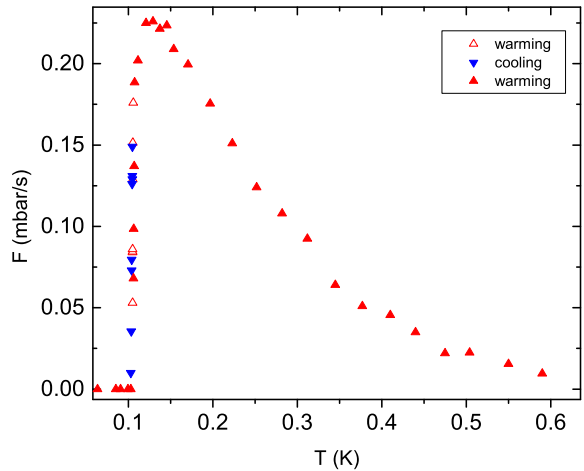}}%
\caption{(color online). Maximum values for the flux as a function of temperature for the case of a solid with $19.5$~ppm $^3$He at $P = 26.40$~bar. The sharp behavior of the flux extinction in a very narrow range of temperature near $T = T_d$ is evident.  \label{Fig-FvsT19ppmdata}}
\end{figure}

The maximum flux measured through the solid helium that results from a specific imposed $\Delta T$ (typically $\pm 10$mK, Fig.~\ref{Fig-10ppm-procedure}) occurs for a resulting $\Delta\mu$ in the range 5-8 mJ/g, and has a temperature dependence as illustrated in Fig.~\ref{Fig-FvsT19ppmdata} for the case of a different solid sample with a \3he concentration of 19.5 ppm. As we will see, this general behavior is present for all of the concentrations we have studied.  This maximum flux, $F$, increases with falling temperature, with warming and cooling showing the same values of the flux for a given sample, so long as the sample is not annealed. And, there is a sharp reversible decrease of the flux at a concentration-dependent temperature, $T_d$.

\begin{figure}[htb]
 \centerline{\includegraphics[width=1.1\linewidth,keepaspectratio]{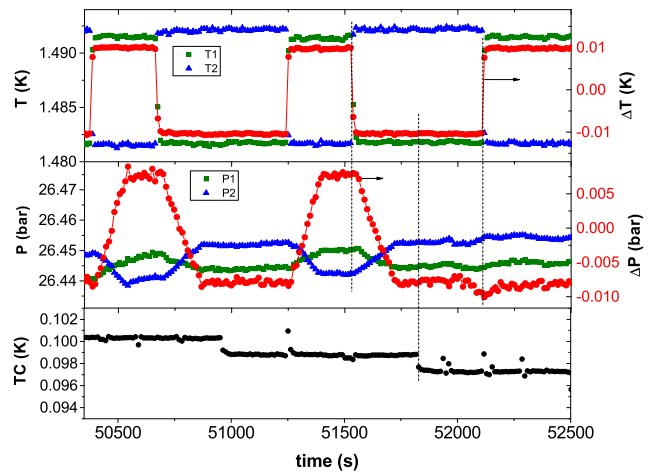}}%
\caption{(color online). An example of the extinction of the flux as the temperature of the solid helium falls below $T = T_d$ for a $10.2$~ppm $^3$He sample at $P = 26.30$~bar. This figure has the corrected temperature scale and is a revision of the similar figure presented in Ref.~\cite{Vekhov2014b}  \label{Fig-10ppm-sharp-cooling}}
\end{figure}

\begin{figure}[htb]
 \centerline{\includegraphics[width=1.1\linewidth,keepaspectratio]{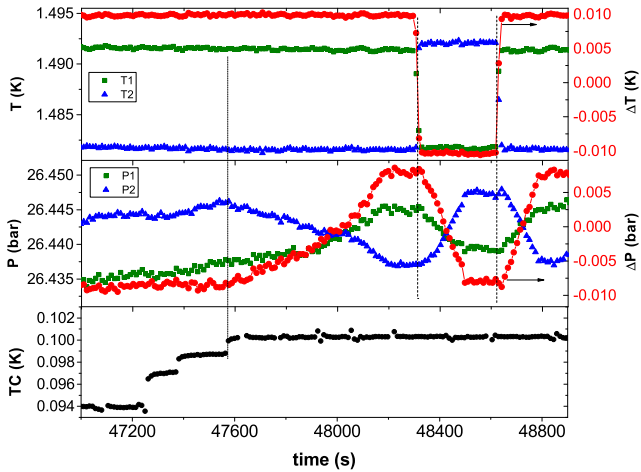}}%
\caption{(color online). An example of the recovery of the flux near $T = T_d$ for the same 10.2 ppm sample as in Fig.~\ref{Fig-10ppm-sharp-cooling}. No flux is seen in the presence of an imposed $\Delta T$ until the cell temperature increases above about 100 mK, after which the flux recovers in a few hundred seconds. This figure has the corrected temperature scale and is a revision of the similar figure presented in Ref.~\cite{Vekhov2014b} \label{Fig-10ppm-sharp-warming}}
\end{figure}

  As seen in Fig.~\ref{Fig-FvsT19ppmdata} for this $\chi$ = 19.5 ppm \3he impurity sample,  the flux that results from a given $\Delta T$ is an increasing function of decreasing temperature until the temperature drops below $\sim$ 105 mK, below which there is no flux. As an illustration of just how sharp and prompt the extinction behavior is, consider Fig.~\ref{Fig-10ppm-sharp-cooling} and Fig.~\ref{Fig-10ppm-sharp-warming}. These figures illustrate that for a $\chi$ = 10.2 ppm sample, the transition from flux to no flux is no more than $\approx$ 1.5 mK wide, with the cessation of the flux complete within no more than $\sim 350$ seconds.  Similarly, in Fig.~\ref{Fig-10ppm-sharp-warming} we see that there is no flow at a cell temperature of 99 mK, but that an increase in the cell temperature to a fixed value near 100 mK results in a growth of the flux, with a flux recovery time of $\sim 600$ seconds. The difference in the temperature of the sharp change in $F$ between cooling and warming shows a small hysteresis at this value of the \3he concentration.  The sharp gradient in the slope of $F$ \emph{vs.} $T$ near 100 mK seen in Fig.~\ref{Fig-FvsT19ppmdata} is stable. That is, if the temperature remains fixed, then the value of $F$ remains stable at any point in the $T_d$ transition region.

Now, the time noted for the flux to make the no-flow to flow recovery (or the reverse) likely places some constraints on scenarios for what causes the transition from a state of no flow to a state of flow. One possibility is that it takes this long for the temperature of the solid to change.  To explore this, a calculation for our cylindrical geometry that incorporates the Kapitza resistance between the solid and the copper wall and the properties of solid helium has been done by Mullin\cite{Mullin-priv-comm} with the result that the time for this thermal equilibration to take place is predicted to be no more than $\sim$~50 msec.  This result is consistent with the thermal equilibration experiments carried out by Huber and Maris\cite{Huber1982} that indicated that equilibration near 100 mK is achieved in $\sim$ 10 msec. These facts indicate that the equilibration time for temperature of the solid is much faster than the observed recovery times and thus the flux change must be related to the movement of the \3he in the solid. We will discuss this further later.

\subsection{Flux Dependence on \3he Concentration}

\begin{figure}[htb]
 \centerline{\includegraphics[width=1.1\linewidth,keepaspectratio]{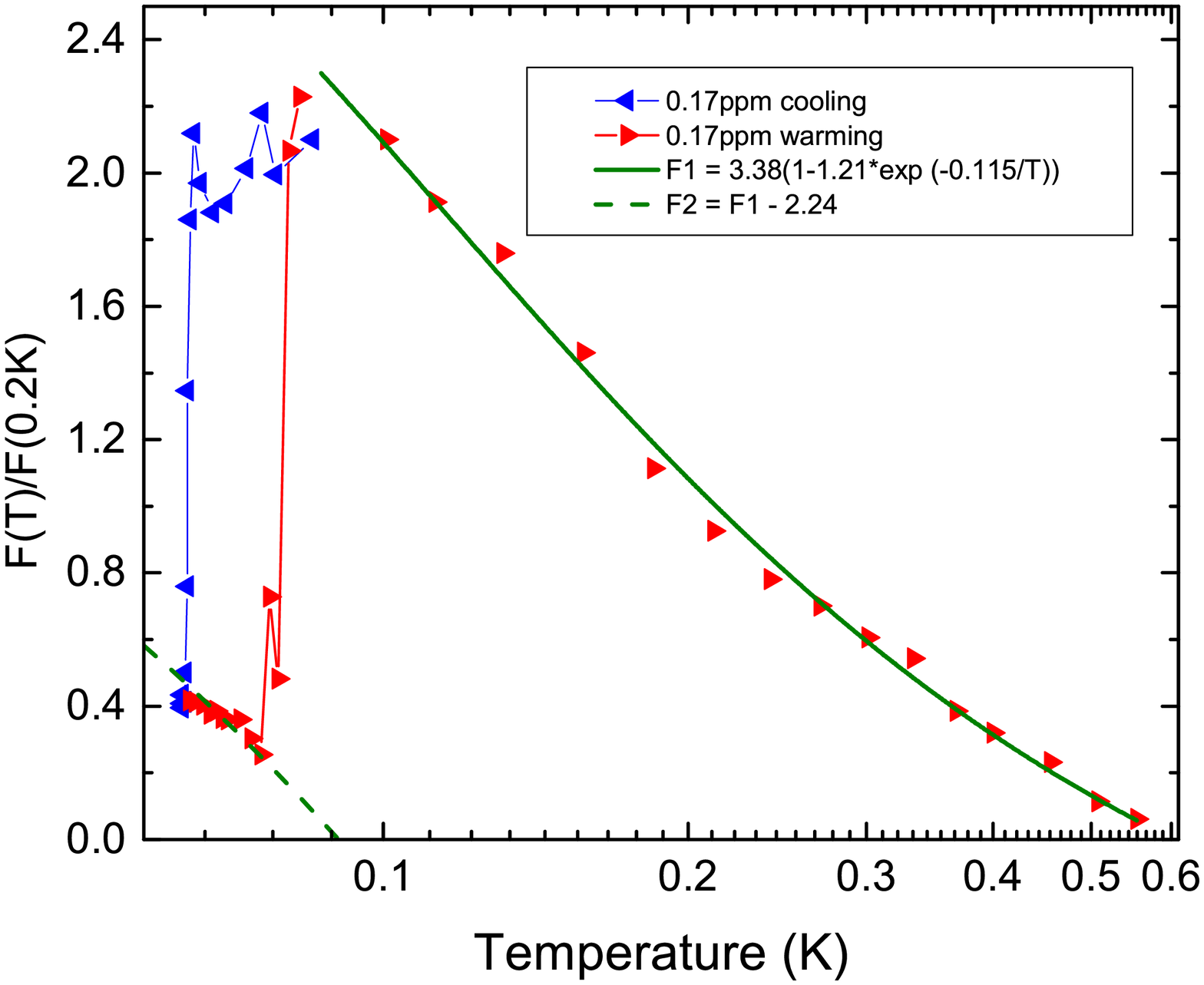}}%
\caption{(color online). An example of flux reduction and hysteresis near $T = T_d$ for a $0.17$~ppm $^3$He  and $P = 25.64$~bar sample. The smooth curve will be discussed in the next section. \label{Fig-FvsT4He}}
\end{figure}

\begin{figure}[htb]
 \centerline{\includegraphics[width=1.1\linewidth,keepaspectratio]{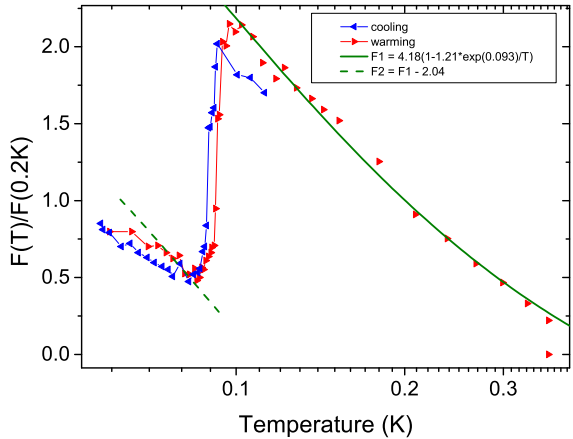}}%
\caption{(color online). An example of flux reduction and hysteresis near $T = T_d$ for a $2.0$~ppm $^3$He  and $P = 26.10$~bar sample. The smooth curve will be discussed in the next section. \label{Fig-FvsT2ppm}}
\end{figure}

The same procedure shown in Fig.~\ref{Fig-10ppm-procedure} has been used for a substantial set of  solid helium samples with different $^3$He impurity concentrations $\chi$ that ranged from a low for nominal well-helium (again, here measured to be 0.17 ppm \3he) to a high of 220 ppm, as listed in Table I. Two examples of the mass flux temperature dependencies are shown in Figs.~\ref{Fig-FvsT4He} and \ref{Fig-FvsT2ppm} for $\chi = 0.17$ and 2~ppm, respectively. Data points here represent the maximum flux values normalized to the maximum flux rate at 200~mK to facilitate the comparison. The relevance of such normalization  will become more apparent shortly.

\begin{table}
\label{tab-samples}
\begin{flushleft}
    \caption{Sample Characteristics}
\end{flushleft}
\centering
\begin{tabular} {l c c c c}
  \hline
  \hline
  &   &   &   &   \\
$\chi$ & $\delta \chi$ & P (bar) & $T_d$ (mK) & $\delta T_d$ (mK)  \\
  &   &   &   &   \\
  \hline
  &   &   &   &   \\
0.17 & -- & 25.64  & 72.5 & 7 \\
0.17 & -- & 25.90  & 72.5 & 3 \\
1.0 & 0.2 & 25.86  & 80.5 & 5 \\
2.0 & 0.2 & 26.10  & 88.5 & 5 \\
4.0 & 0.5 & 26.09  & 91.5 & 5 \\
10.2 & 0.5 & 26.30  & 97.0 & 5 \\
15.0 & 3 & 25.92  & 99.5 & 3 \\
19.5 & 1 & 26.40  & 103 & 3 \\
25.5 & 1.2 & 26.12  & 106 & 3 \\
40.0* & 5 & 26.15  & 109 & 2 \\
61.0 & 3 & 26.36  & 111 & 2 \\
119.3 & 6 & 26.40  & 115.5 & 2 \\
220.0* & 30 & 25.90  & 125 & 2 \\
  &   &   &   &   \\
  \hline
  \hline
\end{tabular}
\\
The $^3$He concentration (ppm) is in two cases estimated (*) based on the $ln(\chi)$ $vs.$ $1/T$ linear dependence shown in Fig.\ref{Fig-Tdvs3He}, inset. The quantities $\delta \chi$ and $\delta T_d$ represent uncertainties in the determination of $\chi$ and $T_d$. 
\end{table}

\begin{figure}[htb]
 \centerline{\includegraphics[width=1.1\linewidth,keepaspectratio]{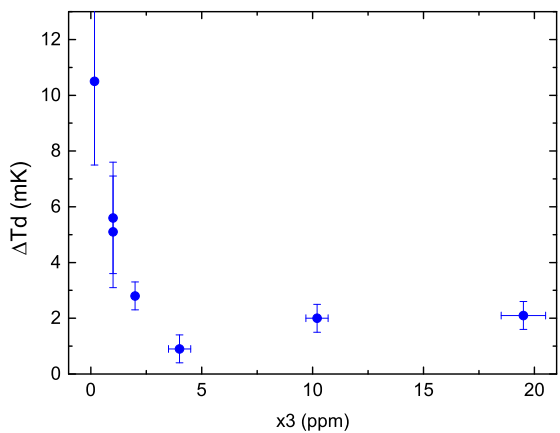}}%
\caption{(color online). The width of the hysteresis seen in the vicinity of $T = T_d$ for low  $^3$He concentrations. The width of the hysteresis region narrows steeply with increasing concentration.  \label{Fig-hysteresis}}
\end{figure}

 Figs.~\ref{Fig-FvsT4He} and \ref{Fig-FvsT2ppm} document hysteresis in the vicinity of $T_d$ with the flux drop during cooling typically taking place at slightly lower temperature than the flux rise during warming. This hysteresis is measurable at most of the $^3$He concentrations studied and might be considered as a feature of a first order phase transition. This suggests that phase separation my be important and we will return to this point later.  The hysteresis is most evident at the lowest concentrations. The width of this hysteresis for low concentrations is shown in Fig.~\ref{Fig-hysteresis}. Figs.~\ref{Fig-FvsT4He} and \ref{Fig-FvsT2ppm} demonstrate that for low concentrations the flux does not drop to zero and recovers on cooling below $T < T_d$. The recovery of the flux as the temperature is lowered below $T_d$\cite{Ray2010c,Vekhov2014b} suggests that the role played by the \3he saturates. We will return to this point in the Comments section, section IV.  The dashed curves on the figures are vertically shifted continuations of the solid smooth curves, which serve to characterize the data.  The significance of these smooth curves will be discussed in the next section. If $\chi$ is more than about 10~ppm, e.g. as shown in Fig.~\ref{Fig-FvsT19ppmdata}, then there is no flux recovery down to $\sim$ 60~mK (the lowest temperature for these measurements).

\begin{figure}[htb]
 \centerline{\includegraphics[width=1.1\linewidth,keepaspectratio]{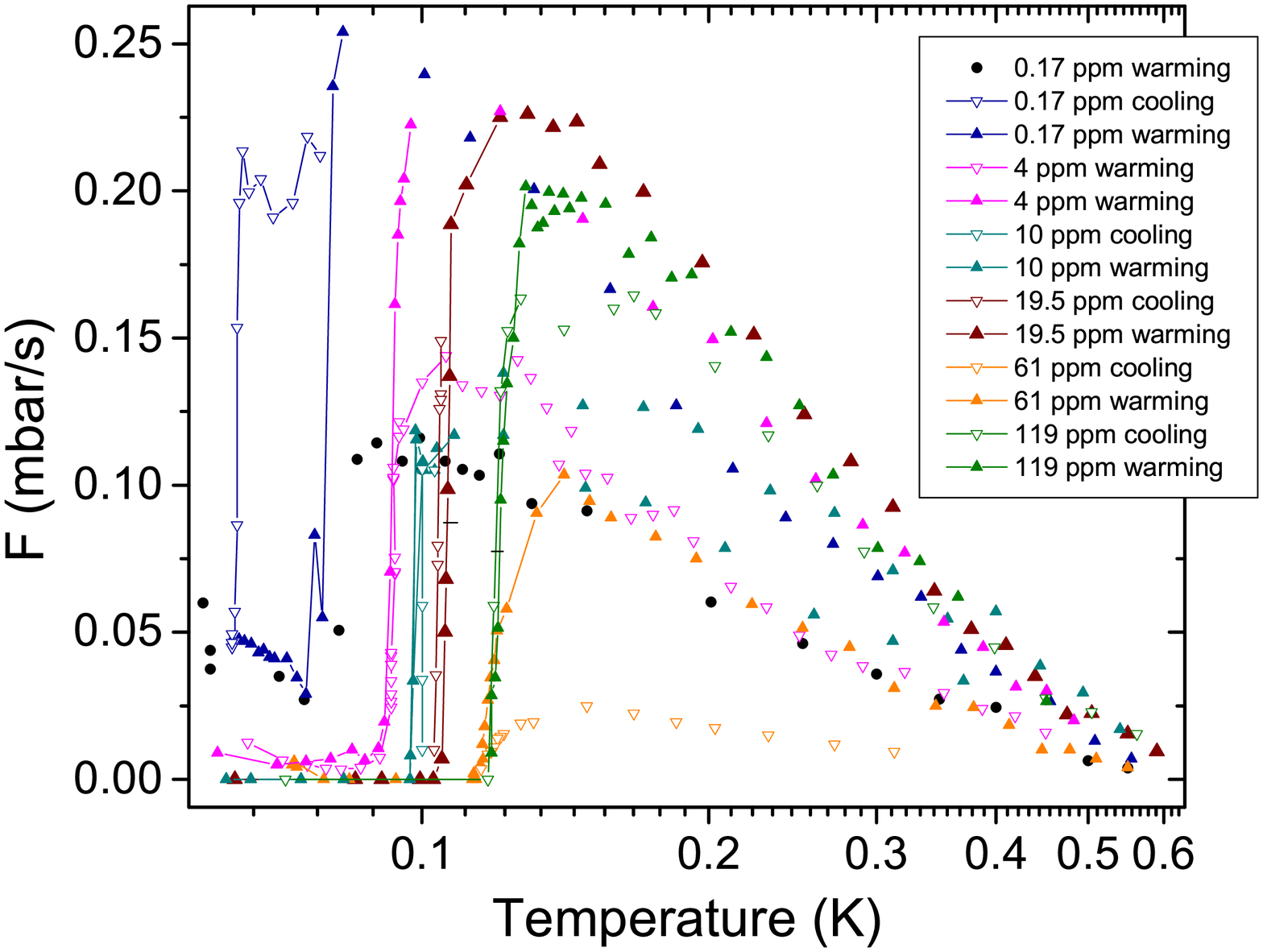}}%
\caption{(color online). The temperature dependence of the
flux observed for $^4$He with several $^3$He impurity
concentrations and experimental conditions, determined in each case with a constant value of $\Delta T$, which yields a maximum flux, $F$, that appears for $\Delta \mu$ in the range 5-8 mJ/g.
This figure has the corrected temperature scale and is a revision of the similar figure, Fig. 1, presented in Ref.~\cite{Vekhov2014b}. For each data set the solid pressure was in the range 26 $\pm$ 0.4 bar.
\label{Fig-non}}
\end{figure}

 Data for a range of samples with different concentrations and sample histories are shown in Fig.~\ref{Fig-non}. In each case the maximum flux value shown is that which results from the same value of the imposed $\Delta T$.  The shift in $T_d$ with concentration is evident.  Different samples with different histories at a given concentration have somewhat different absolute values of $F$, but the temperature dependence and value of $T_d$ are reproducible for a given concentration.

\begin{figure}[htb]
 \centerline{\includegraphics[width=1.1\linewidth,keepaspectratio]{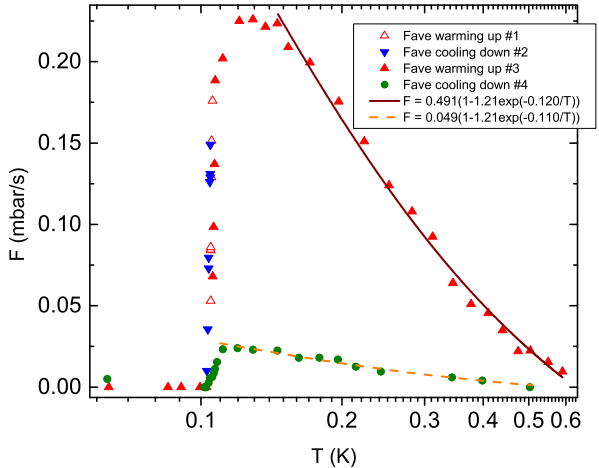}}%
\caption{(color online). An example of flux extinction near $T = T_d$ for a $19.5$~ppm $^3$He sample at $P = 26.40$~bar sample. For comparison, data for cooling is shown (solid circles) after the flux ceased at $T = 0.625$~K. Note, flux values are shown in mbar/s units.  \label{Fig-FvsT19ppm2}}
\end{figure}

\section{DISCUSSION}
\subsection{Universal Temperature Dependence}
We now present in more detail the temperature dependence for temperatures above $T_d$.  As we will show, the temperature dependence for $T > T_d$ is robust, but the absolute value of the flux depends importantly on the sample and its history. As an example of the sort of variability that we have found consider the data shown in Fig.~\ref{Fig-FvsT19ppm2} for the sample with $\chi = 19.5$~ppm. The flux values are presented here in mbar/s units (not normalized values) in order to compare the behavior of the flux before and after the temperature was increased to 620 mK,   where the flux was no longer measurable. When cooled, after the sample cell was warmed, the flux was greatly reduced (circles). The data set for the larger values of $F$ shown here is the same data set shown in Fig.~\ref{Fig-FvsT19ppmdata}.

\begin{figure}[htb]
 \centerline{\includegraphics[width=1.1\linewidth,keepaspectratio]{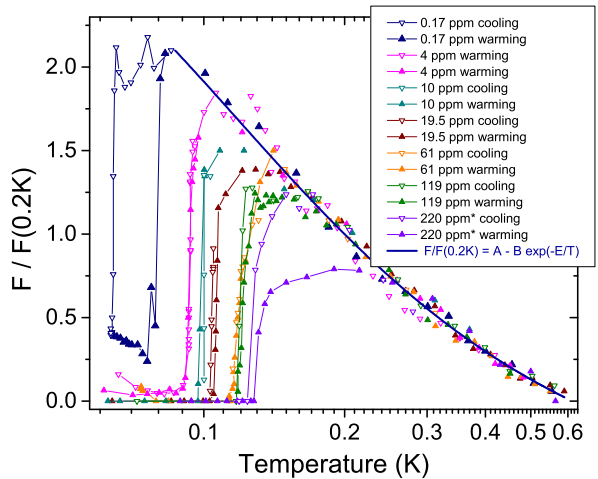}}%
\caption{(color online). The temperature dependence of the
normalized flux observed for $^4$He with several $^3$He impurity
concentrations and experimental conditions, with the solid \4he pressure $26 \pm 0.2$ bar. Fitted line: see text. \label{Fig-FvsTall}}
\end{figure}

What is not immediately apparent in  Fig.~\ref{Fig-non}  and Fig.~\ref{Fig-FvsT19ppm2} is that the temperature dependence at temperatures above the peak flux reached is robust.  To demonstrate this most clearly, we normalize the many data sets shown in Fig.~\ref{Fig-non}. We accomplish this by use of a multiplicative factor for each data set to force the various values of $F$ to superimpose at $T$ = 0.2~K.  The normalized flux temperature dependencies for the samples of different $^3$He concentrations are presented in Fig.~\ref{Fig-FvsTall}.

One can see again here, as was evident in Fig.~\ref{Fig-non}, that the $T_d$ values shift to higher temperatures with higher $\chi$ values. At the same time, $F(T)$ for different samples at $T > T_d$ collapse to a universal temperature dependence \cite{Vekhov2014b}. This Figure also shows that in the temperature range in the vicinity of the peak value of the flux (near $T = T_d$) the peak becomes more rounded with less curvature for larger $\chi$ values, as does the slope of $T_d$ \emph{vs.} $\chi$. 

\begin{figure}[htb]
 \centerline{\includegraphics[width=1.1\linewidth,keepaspectratio]{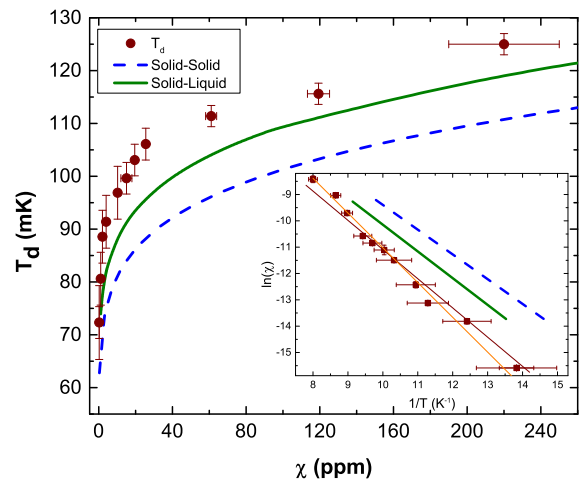}}%
\caption{(color online). Temperature of the sharp drop in $F$, $T_d$. Inset: $ln(\chi)$ $vs.$ $1/T$; see text. This figure has the corrected temperature scale and is a revision of the similar figure, Fig. 5, presented in Ref.~\cite{Vekhov2014b}.
\label{Fig-Tdvs3He}}
\end{figure}

$T_d$ values determined from the data in  Fig.~\ref{Fig-FvsTall} and other data like them are shown in Fig.~\ref{Fig-Tdvs3He} to show how the  $^3$He concentration affects $T_d$. It is natural to compare these temperature-dependent data to the data on phase separation in solid helium. According to Ref.~\cite{Edwards1989}, extrapolated to 26 bar, the temperature of Solid-Solid (bcc $^3$He-rich inclusions form inside the hcp $^4$He-rich matrix) phase separation temperature, $T_{ps}$, can be found from

\begin{equation}
T_{ps} = \left[0.80(1-2\chi)+0.14\right]/\ln(1/\chi-1)
\label{Eq-Tps}
\end{equation}
\\
and this is represented in Fig.~\ref{Fig-Tdvs3He} by the dashed line. The number $0.80$ in this expression comes from the extrapolation to the pressure of our experiment. In the case of bulk phase separation for our pressure range, another situation is present: liquid $^3$He-rich regions form inside the solid \4he matrix (the so-called, Solid-Liquid case). This scenario was calculated in Ref.\cite{Vekhov2014b} and is shown by the solid line in the same Figure. It can be seen here that our $T_d$ temperatures lie above the Solid-Liquid case. This will be discussed further in Section~\ref{discussion}.

\begin{figure}[htb]
 \centerline{\includegraphics[width=1.1\linewidth,keepaspectratio]{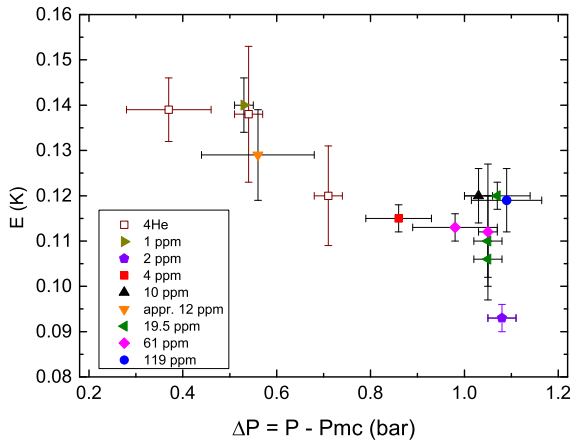}}
\caption{(color online). Pressure dependence of the parameter $E$.  \label{Fig-EvsdP}}
\end{figure}

In an attempt to further characterize the data we have utilized several functions. In recent presentations\cite{Vekhov2012,Vekhov2014b}, we have favored $F = A - B\exp(-E/T)$, which is motivated by the thought that some thermally activated process may be relevant.  Independent fits of this functional dependence  to all of the data sets results in the recognition of the universal character of the temperature dependence. We find a good characterization of the data with .

\begin{equation}
F = F_0[1 - 1.21\exp(-E/T)],
\label{expfit}
\end{equation}
\\
  The value of $E$ that results from such a characterization of the data depends weakly on pressure as shown in  Fig.\ref{Fig-EvsdP}.  The higher the pressure (density), the lower the value of $E$, i.e. the $F(T)$ dependence gets steeper with pressure. The use of colors for symbols in this Figure is the same as in Fig.~\ref{Fig-FvsT19ppm2} and Fig.~\ref{Fig-FvsTall}.  There is no apparent dependence of $E$ on $\chi$.

The value of $F_0$ may be interpreted to be proportional to the number of conducting pathways inside the solid helium. I.e. at $T \sim T_h$ conducting pathways are being partially annealed (completely in some cases) leading to a substantial flux decrease in the whole $T_d < T < T_h$ temperature range on subsequent cooling.

As we have pointed out earlier\cite{Vekhov2014} the entire temperature dependence is not fully explained by thermal activation since the flux extrapolates to zero at a finite temperature. So, whatever controls the decrease in flux with increasing temperature must have an explanation that goes beyond simple thermal activation.

\begin{figure}[htb]
 \centerline{\includegraphics[width=1.1\linewidth,keepaspectratio]{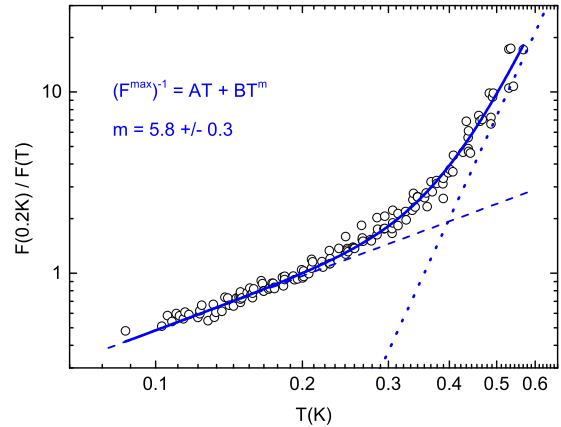}}%
\caption{(Color online) Temperature dependence of the flux resistance, $(F)^{-1}$, measured through the solid sample (see Ref.\cite{Vekhov2014b}, Fig.4) and presented here on log-log scale. The solid line is a fit of the data by Eq.~(\ref{Eq_1FCvslogT}) and the dashed and dotted lines represent linear and $T^{5.8}$ behavior, respectively.}
 \label{Fig-1FCvslogT-old}
\end{figure}

 As an alternate approach to characterize the temperature dependence, the normalized universal data for $T > T_d$ from Fig.~\ref{Fig-FvsTall} can be inverted, $(F)^{-1}$,  to obtain something we might call a flux resistance as shown in Fig.~\ref{Fig-1FCvslogT-old}. This approach allows us to explore whether there might be any power law behavior, although the temperature range is very narrow for such an approach.  One can see in Fig.~\ref{Fig-1FCvslogT-old} that there appears to be a crossover in the behavior of the temperature dependence. The data for the range of samples studied can be described reasonably well by 
\begin{equation}
(F)^{-1} = F(0.2K)/F = AT^k + BT^m,
\label{Eq_1FCvslogT}
\end{equation}\\
where $A$, $B$, $k$ and $m$ are parameters. We find that with the choice of $k = 1$ a fit to the data yields $m = 5.8 \pm 0.3$.  

It is an open question as to what the origin of those two apparently distinct contributions to the temperature dependence of the mass flux resistance is. But, such behavior is not without precedent for a quasi-one-dimensional system.
Consider, for example, the case of superconducting nanowires below the transition temperature, $T_c$. These nanowires demonstrate nonzero resistance at any finite temperature, apparently due to the presence of phase slips in the order parameter that result in dissipation, which destroys superconductivity. These phase slips are due to thermal fluctuations at higher temperatures close to, but below $T_c$, or to quantum-mechanical tunneling at low temperatures, so-called quantum phase slips (QPS).
Electrical transport measurements in single-crystal Sn nanowires \cite{Tian2005} and its analysis \cite{Pai2008} showed a power-law dependence $\rho(T) \sim T^{\alpha}$ at $T < T_c$ with an exponent $\alpha \approx 5$ for nanowires of diameter 20 and 40~nm, but much larger values of $\alpha$ for larger wire diameters. These data were interpreted in terms of the unbinding of quantum phase slips with temperature. Also it was predicted \cite{Zaikin2008} that in the limit of very thin wires and low temperatures, where unbound QPS behave as a gas, the temperature dependence of the wire resistivity should become linear at the transition to the disordered (i.e. nonsuperconducting) phase. As shown in Fig.~\ref{Fig-1FCvslogT-old}, the data is consistent with a linear dependence of the flux resistance for the flux that we observe at low temperature. The relationship that might exist between these rather different physical systems, conducting pathways in solid helium and very thin wires, has not been explored theoretically, but both systems may be describable by Luttinger liquid theory.

\subsection{Luttinger Liquid}
\label{Luttinger}

\begin{figure}[htb]
 \centerline{\includegraphics[width=1.1\linewidth,keepaspectratio]{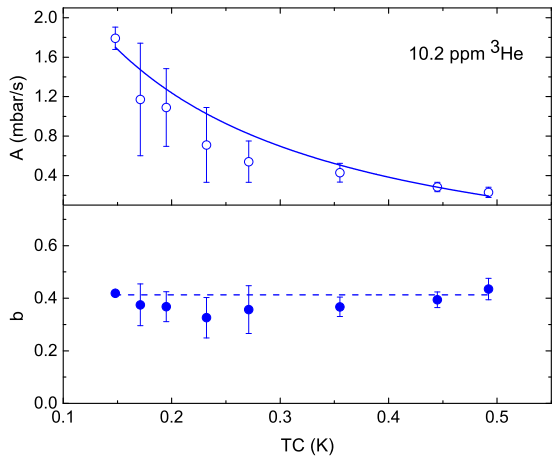}}%
\caption{(color online). Temperature dependence of fit parameters A (a) and b (b) for the data in Fig.\ref{Fig-10ppm-Fvsdmu}; see Eq.(\ref{Eq-Fvsdmu}). \label{Fig-AbvsT}}
\end{figure}

The non-linear behavior of  $F$ vs. $\Delta \mu$ shown in Fig.~\ref{Fig-10ppm-Fvsdmu} is reasonably well represented by Eq. (3), where $b$ is less than 0.5, independent of $T$.  An example of the independence of $b$ from temperature is shown in Fig.~\ref{Fig-AbvsT} for $\chi$ = 10.2 ppm. Here, as in our earlier measurements for nominal purity well helium\cite{Vekhov2012}, all of the temperature dependence is contained in the amplitude $A$. We have previously shown that for well helium $b$ depends on pressure. As we have suggested, this non-linear behavior and independence of $b$ from temperature supports the possibility that the flux is carried by one-dimensional paths, e.g. perhaps the cores of edge dislocations \cite{Vekhov2012}, and can be described by the properties of a, so-called, Bosonic Luttinger liquid \cite{Luttinger1963}. 

\begin{figure}[htb]
 \centerline{\includegraphics[width=1.1\linewidth,keepaspectratio]{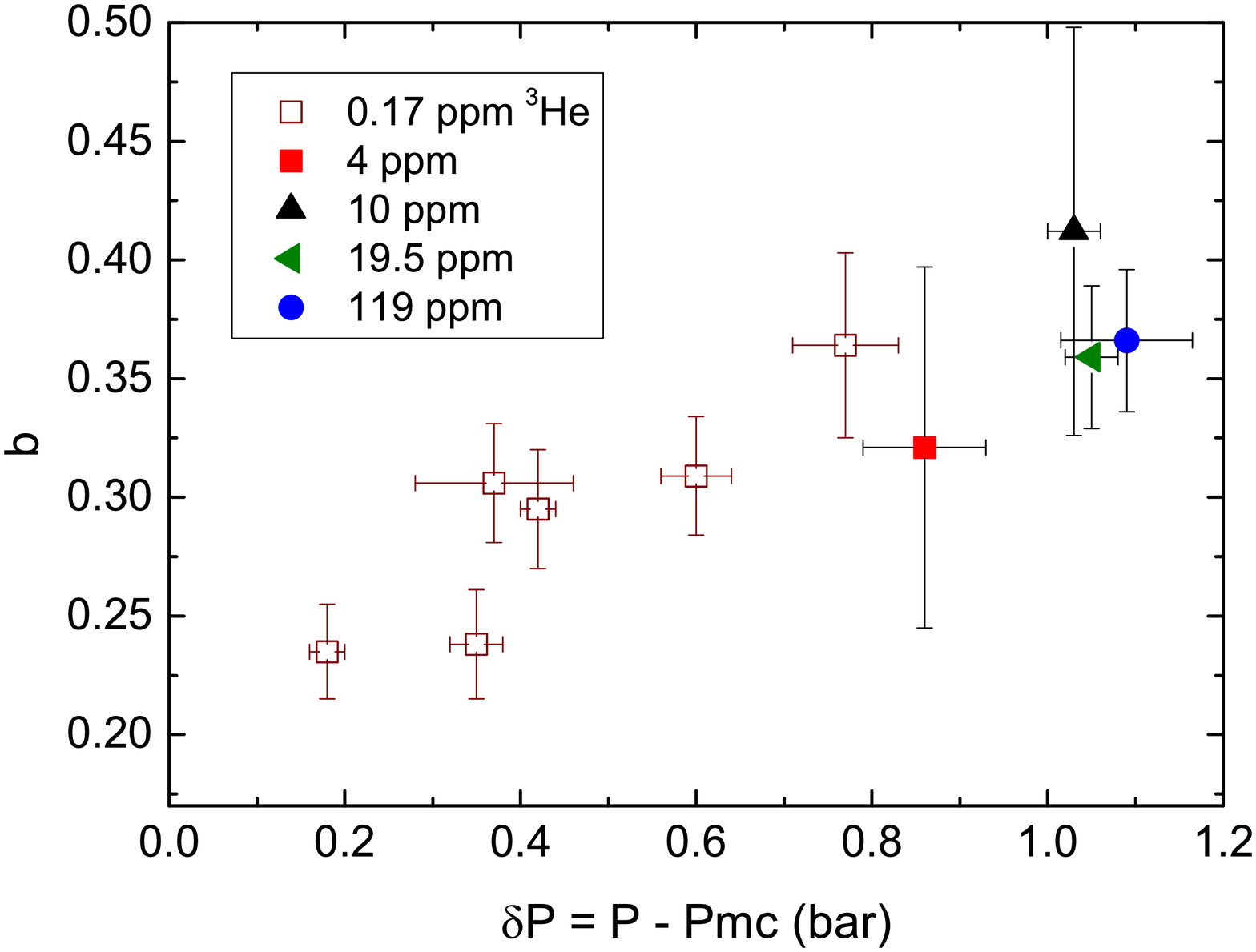}}
\caption{(color online). Pressure dependence of the fit parameter $b$. Open square data points here correspond to nominally pure $^4$He samples (0.17~ppm $^3$He) and other data points represent the data for $^3$He-$^4$He mixtures with $\chi > 0.17$~ppm. Here, $\delta P$ is the distance above the melting curve in bar. \label{Fig-bvsdP}}
\end{figure}

As noted, the exponent $b$ is temperature independent, but it depends on the solid helium pressure. The higher the pressure, the larger is the value of $b$. Data for $b$ as a function of the distance from the melting curve, $\delta P = P - P_{MC}$ is presented in Fig.~\ref{Fig-bvsdP} for nominally pure (170 ppb) helium as well as for a number of concentrations. Although the number of concentrations for which we have data adequate to determine $b$ for a range of pressures is limited, there is apparently no significant dependence of $b$ on $\chi$.

\begin{figure}[htb]
 \centerline{\includegraphics[width=1.1\linewidth,keepaspectratio]{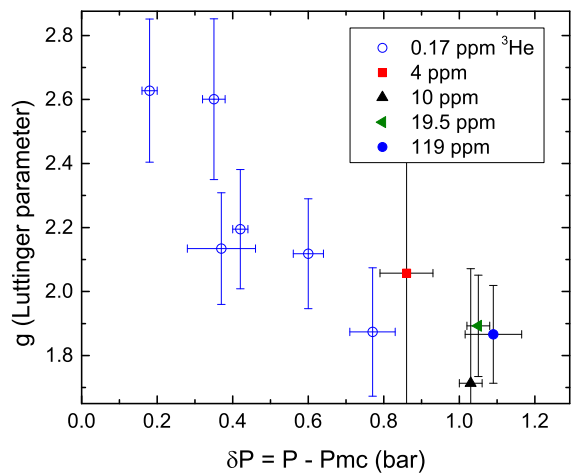}}
\caption{(color online). Pressure dependence of the Luttinger
parameter, $g$, presuming that $g = [(1/b) + 1]/2$. Again here, $\delta P$ is the distance above the melting curve in bar. \label{Fig-gvsdP-new-20}}
\end{figure}

The data for $b$ above can be used to obtain the pressure dependence of the Luttinger parameter, $g$. If we presume that we have a number of independent random scattering sites that introduce phase slips, then the Luttinger parameter $g$ can be obtained from $b$ by means of $g = [(1/b) + 1]/2$ \cite{Svistunov2012}. The results shown in  Fig.~\ref{Fig-gvsdP-new-20} suggest that for such a scenario we are well in the Luttinger regime, but that with increasing pressure $g$ decreases and we approach the non-superfluid regime. This is consistent with previous work\cite{Ray2008a,Ray2009b} in which the flux disappeared at higher pressures.   The colors of symbols in Fig.~\ref{Fig-gvsdP-new-20} have the same sense as in Fig.~\ref{Fig-bvsdP}.

Based on the pressure (density) dependence of $g$ and $E$, Figs.\ref{Fig-gvsdP-new-20} and \ref{Fig-EvsdP}, respectively, one can see that $^3$He impurities in its range studied do not affect these data. This suggests that for $T > T_d$ there is no measurable $^3$He role in either the $F(\Delta \mu)$ or the  $F(T)$ dependencies.

\subsection{COMMENTS}
\label{discussion}

We have discussed the effect of $^3$He impurities on the flux measured and have shown that $T_{ps} \neq T_d$.
We summarize the $^3$He effects that we have observed. First of all, there is a sharp flux extinction at a characteristic temperature $T_d$, which itself depends on $^3$He concentration, $\chi$, and this $T_d(\chi)$ dependence (see Fig.\ref{Fig-Tdvs3He}) is reminiscent of the bulk phase separation shifted to higher temperatures. Another point is that this extinction  is a slow process compared with the thermal equilibration time for the solid ($\sim$~10's of msec, but a fast process ($\sim$~100's of seconds), compared to the time required for complete solid phase separation (dozens of hours \cite{Ganshin1999}) in samples of higher concentrations than we have used here.  These effects could suggest that only a small fraction of $^3$He is responsible for the flux extinction and this phenomenon is due to $^3$He redistribution in the vicinity of the phase separation temperature. Work by Edwards et al.\cite{Edwards.1962} indicates deviations from $T^3$ behavior in the specific heat for $T > T_{ps}$, which suggests local \3he concentration fluctuations\cite{Antsygina.2005} are likely relevant to our observations; the \3he fluctuates in position and is thereby able to block the flux at a temperature above the bulk phase separation temperature. It is the case that for the very low concentrations we have for the most part used in this work, diffusion can be quite fast. We will explore this further below.

Based on the $F \sim (\Delta \mu)^b$ dependence, as was already shown in Ref.\cite{Vekhov2012}, we suggested that the flux could be consistent with a one-dimensional scenario. One can envision two possible candidates for these 1D pathways: (1) liquid channels, e.g.  between grain boundaries and the sample cell wall\cite{Sasaki2008} or  (2) the cores of dislocations in solid helium.

\subsubsection{Liquid Channels}

 Although we have previously argued that liquid channels are likely not present, we discuss them further here.  Were they present, the superfluid density in them could be reduced or eliminated by the migration of \3he.  Unfortunately, there is no reasonable estimate of how many of these liquid channels there might be.  In the pressure range we have studied, the diameter of these channels can be calculated \cite{Sasaki2008} to be $6-31$~nm; the higher the pressure, the smaller the channel diameter. In the most extreme example, if one such liquid channel were to span the distance between the Vyror rods with diameter of 20 nm, to fill it with \3he would require $1.4 \times 10^{11}$ atoms. At 10 ppm \3he concentration there are enough \3he atoms present in our solid mixture samaple to fill $\sim 3 \times 10^{6}$ such channels. Although we don't have a good estimate of the number of such channels, given the expected diffusion times it appears that diffusion to such channels would be a fast process.

\begin{figure}[htb]
 \centerline{\includegraphics[width=1.1\linewidth,keepaspectratio]{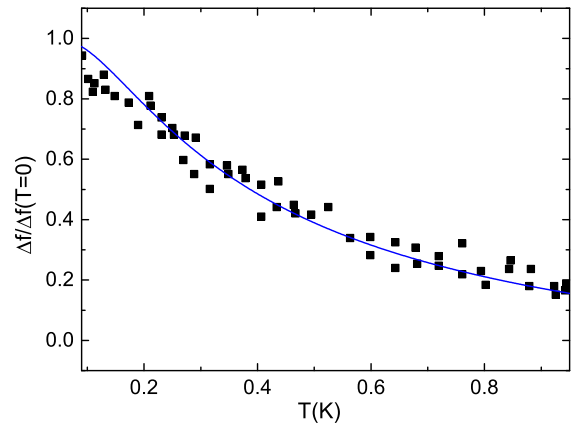}}
\caption{(color online). Temperature dependence of the torsional oscillator data from Ref.\cite{Ikegami2007} for a pore diameter of 1.8 nm. Here $\Delta f/\Delta f(T=0)$ is the relative frequency shift seen in the torsional oscillator as a function of temperature. The smooth curve is a fit of the data directly to Eq.~(\ref{expfit}) that we have used to characterize the universal temperature dependence of our flux through solid helium; we find in this case $E \sim 0.4$ K. If we convert to a flux resistance in this case, the power law, Eq.~(\ref{Eq_1FCvslogT}), does not provide a reasonable fit. \label{ikegami}}
\end{figure}

 In work reported in Ref. \cite{Ikegami2007}, torsional oscillator, TO, measurements to determine the superfluid density, $\rho_s$, were carried out for nanometer-size channels (folded sheet mesoporous materials) of diameter $D = 1.5 - 4.7$~nm \cite{Ikegami2007, Taniguchi2010} filled with superfluid helium. This study revealed a transition from a Kosterlitz-Thouless behavior to a 1D-like temperature dependence of the apparent superfluid density only for $D < 2.2$~nm. The temperature dependence they found, $\rho_s$ for $D = 1.8$~nm \cite{Ikegami2007},
can be well fit by Eq.(\ref{expfit}) which has been chosen to fit our flux temperature dependence, $F(T)$ (see Fig.~\ref{ikegami}). A value of $E \sim 0.4$~K for the parameter $E$ is found for the data of Ref.\cite{Ikegami2007}. This functional dependence, which is not present in the work of Ref.\cite{Ikegami2007} above $D$ = 2 nm, lends support to the notion of a 1D scenario for our observed flux and further suggests that the liquid channels that are predicted would be too large in diameter to demonstrate 1-D behavior.

\subsubsection{Dislocations}

Based on the possibility of 1D of conducting pathways\cite{Soyler2009}, one approach is to assume that $^3$He impurities bind on dislocation cores  (or their intersections) in the solid helium and block the flux that is carried by such pathways. This notion is supported by QMC simulations \cite{Corboz2008} that show that $^3$He impurities diminish the superfluid density along the core of screw dislocations in hcp $^4$He by binding on them.
To illustrate this point, the inset to Fig.\ref{Fig-Tdvs3He} shows $\ln(\chi)$ vs $1/T$. Straight lines here are presented by $\chi = \exp(-R/T)$ with $R$ approximately independent of temperature, and $R=0.94$~K and 1.02~K for Solid-Solid (dashed curve) and Solid-Liquid (solid curve) bulk phase separation, respectively\cite{Vekhov2014b}. A fit of the $T_d$ data (squares, dark solid line) by $\chi = \exp(-R/T)$ gives $R = 1.11$~K. A model\cite{Vekhov2014b} that includes a small number of binding sites for $^3$He or $^4$He atoms yields the function form $\chi = \exp(a-R/T)$, where $\exp(a)/(1+\exp(a))$ is the minimum impurity concentration that blocks superflux, and $R$ includes the binding energy. This function form gives better fit (solid red line), with $R = 1.32$~K and $a = 2.18$. The numbers reported in this paragraph are revisions of those previously reported\cite{Vekhov2014b} because they take into account the temperature scale revision mentioned earlier in this report.  This energy value is higher than the predicted \cite{Kim2011,Corboz2008} binding energy ($\sim 0.7$~K) of single $^3$He atoms to dislocation cores. Although to our knowledge it has not been calculated, the binding energy to dislocation intersections should exceed this.  These facts are consistent with the possibility that the flux extinction results from the $^3$He binding to dislocation intersections \cite{Corboz2008}, where the $^3$He blocks the flux.  To our knowledge, the binding energy for \3he at the intersections of dislocations has not been considered theoretically and is not known.

It is perhaps useful to carry out quantitative estimates that relate to the decoration of such cores (or intersections) with \3he atoms. There are a number of unknowns. One of these is the number of such structures that span the cell, Vycor to Vycor. We previously took\cite{Vekhov2012} this number to be of order $10^5$.  If we take the solid \4he density to be $\approx 2.9 \times 10^{22}$ atoms/cm$^3$ we find that about $6 \times 10^7$ \4he atoms would be along a 2 cm direct strictly 1-D pathway between the two Vycor rods; a pathway of diameter $\sim$ 1 nm\cite{Soyler2009} would require $\sim 5 \times 10^8$ atoms. Thus to fully decorate $10^5$ such cores would require $\sim 10^{13}$ atoms (or $\sim 10^{14}$ for a 1 nm diameter case).  Since Corboz et al.\cite{Corboz2008} have shown that the decoration of dislocation cores does not have to be complete to influence the superfluidity on the core, we rather arbitrarily take a smaller number, $\sim 1 \times 10^{12}$ atoms. Of course, a much smaller number could also be relevant since only a local dense decoration would be needed along a short segment of a dislocation core or at an intersection to substantially reduce the flux. None the less, we continue with this number.  Now, in our experimental geometry a \3he concentration of 10 ppm results in the presence of about $5.4 \times 10^{17}$ \3he atoms in the cell, which, if uniformly distributed is a \3he density of $2.9 \times 10^{17}$ atoms/cm$^3$.

So, we can ask from what volume the needed number of atoms (e.g. for the $10^{13}$ case) would have to diffuse in the solid to decorate the dislocation cores. If we assume that these atoms diffuse from a cylindrical region to a dislocation core on the axis of the cylinder we can estimate the radial distance from the core over which they would have to travel.  We find this radial distance to be $\sim 1.5 \times 10^{-4}$ cm.  We can then ask how long this will take. Those from a region near the core will arrive relatively quickly while those from further away (an increasing number in any radial interval) will arrive later. For this we take the \3he diffusion constant near 26 bar and 100 mK at $\sim$ 10 ppm (extrapolating from the work of Eselson et al.\cite{Eselson1978}) to be $D \approx 1.5 \times 10^{-6}$ cm$^2$/sec and $<x^2> \sim 6Dt$, where $t$ is the time, and we find the time to be about $2.5 \times 10^{-3}$ seconds.    For higher \3he concentrations the atoms would have to travel less far, but for higher concentrations the diffusion constant is smaller.   The time for flux recovery at $T_d$ is well documented in Fig.~\ref{Fig-10ppm-sharp-warming}, but the time for flux extinction is not yet well determined; we can only say that it is apparently less than $\sim 200$ seconds.  In spite of the approximations involved, if we compare the computed numbers for diffusion times with the times cited near Fig.~\ref{Fig-10ppm-sharp-cooling} and Fig.~\ref{Fig-10ppm-sharp-warming} for the flux to be extinguished or recover, we conclude that diffusion to dislocations or their intersections is not likely a limiting factor in the flux change at $T_d$; it would likely be a rather fast process for the concentrations that we have studied.

\subsubsection{A Helium Film}

Another unlikely possibility is the flow of a helium film along the surfaces of the Vycor and the experimental cell.  Were this to be the cause of the mass flux, one might expect that the temperature dependence would behave like a Kosterlitz-Thouless transition.   This behavior is not evident in the data shown in Fig.~\ref{Fig-FvsT19ppmdata} or other similar sets of data. None the less, were we to imagine a surface layer of liquid \4he at the walls with a thickness of two atomic layers (above a solid-like layer adjacent to most surfaces; above that layer \4he is known to be located adjacent to a wall in liquid mixture situations),
the number of \4he atoms involved would be $\sim 1.4 \times 10^{16}$. At 1 ppm \3he this is approximately the number of \3he atoms in the cell. Given diffusion times, it is conceivable such a film could be poisoned.  But, the fact that annealing reduces or eliminates the ability of the solid to carry a flux when subsequently cooled argues strongly against a superfluid film as the carrier of the flux.

\subsubsection{Vycor Pore Openings}

Another possibility for what causes a reduction in the flux at $T_d$ is \3he accumulation at the openings to the Vycor pores. This possibility has recently been emphasized by Cheng et al.\cite{Cheng2015}. In their experiments a variation of our approach was used. Instead of a superfluid-solid-superfluid geometry, they used a solid-superfluid-solid geometry.  They were able to observe some temperature dependencies that are similar to those we have found in our various experiments, particularly the presence of $T_d$.  In their discussion of the pore opening scenario, which they supported by calculations of the temperature-dependent binding of \3he to various possible binding sites, the picture is that at $T_d$ the \3he moves from the solid to the interfaces where the solid helium in the cell meets the liquid helium at the openings of the Vycor pores.

For the flux to be fully blocked by \3he in this scenario, the openings of the pores where the superfluid in the Vycor meets the solid must be blocked by \3he. For our experimental apparatus each Vycor rod surface meets the solid \4he over a macroscopic surface area of about 0.3 cm$^2$. Given the properties of Vycor, we estimate that the open area of this surface that is comprised of pore openings is no less than $0.084$ cm$^2$.  Each pore of diameter 7 nm will have an open area at the surface of the Vycor no less than $3.84 \times 10^{-13}$ cm$^2$. The number of such pore openings is estimated to be at least $2.18 \times 10^{11}$. All of these need to be blocked by \3he. How much is needed at each pore opening?  It is likely that 1 monolayer will not be adequate.  We take as an estimate a distance along the pore of two pore diameters and presume that if this volume were to fill with \3he the pore would be blocked; a greater length certainly might be required. In the vicinity of 100 mK the expected phase separation for liquid \3he indicates that if the \3he were to be in the pores and blocking the flux it would be due to a high concentration normal mixture in the pores. Clearly our numbers provide only a rough estimate, but it is not unreasonable.

To full such a volume for all of the pores would require $\approx 2.6 \times 10^{15}$ atoms. At 10 ppm, this is about 0.5 percent of all of the \3he available; for higher (lower) concentrations it is a proportionally smaller (larger) fraction.  So, we can ask in this case about how long it will take for the \3he to accumulate at the pore openings. We take the same parameters of 26 bar, 10 ppm \3he with a diffusion constant of $\approx 1.5 \times 10^{-6}$ cm$^2$/sec and note that the \3he will have to travel macroscopic distances. For the case of 10 ppm \3he we find that the time required is $\sim 20$ seconds.  Changing the concentration changes the time required; increasing the length of the pore that needs to be filled with \3he increases it. These estimates are in very rough order of magnitude agreement with the times shown for recovery of the flux documented in Fig.~\ref{Fig-10ppm-sharp-warming}, especially when it is recognized that as \3he accumulates in the solid near a pore opening the local concentration may increase, which will cause the diffusion constant to decrease\cite{Eselson1978}.  As we have noted, flux extinction is complete in  $\sim 200$ seconds; recovery in $\sim 500$ seconds.

To explore the flux increase for $T < T_d$ at the lowest concentrations it may be that the \3he is exhausted and incompletely effective in blocking the pores. For the case of 0.17 ppm \3he the filled cell will have $9.2 \times 10^{15}$ \3he atoms in it. As we have noted, to fill the pores to a depth of two pore diameters will require $\approx 2.6 \times 10^{15}$ atoms. We have seen that the flux recovers for \3he concentrations up to 10 ppm (in which case there are $5.4 \times 10^{17}$ \3he atoms in the cell).  These estimates have a number of assumptions; in spite of them it is not clear why there is not enough \3he to completely block the flux. The transition to low flux is prompt, but for low concentrations is not complete. One probable cause for this is that at a given temperature which is $T_d$ for the concentration in the solid, once \3he atoms begin to leave the solid matrix, the solid now contains a lower concentration of \3he. This naturally shifts the $T_d$ to a lower temperature and at the given temperature no additional \3he atoms leave the matrix. This observation explains the fact clearly seen in, for example, Fig.~\ref{Fig-FvsT19ppmdata} and evident in other data sets, that the flux in the middle of the $T_d$ region is stable if the temperature is fixed. That is, the data taken while cooling or warming is stable once a fixed temperature is achieved.

\subsubsection{Behavior above and below  $T_d$}

 It is not clear what might explain the universal temperature dependence above $T_d$. And, unless the estimates made here are substantially in error, it is also not fully clear why the flux increases for low \3he concentrations below $T_d$.  One possibility below $T_d$ is that if the \3he has been exhausted, but the blockage incomplete, then a lower temperature could be expected to provide an increase in the superfluid density in the confining conductance pathway.  In very recent work we have reported evidence that for the region of this universal temperature dependence, $T > T_d$, the limitation to the flux is unrelated to the Vycor interface with the solid and takes place in the bulk solid\cite{Vekhov2015R}.  This suggests that this temperature dependence may be due to a temperature-dependent superfluid density along the conducting pathways.  Additional work is currently in progress and will be discussed more extensively in a future publication.






\section{CONCLUSIONS}

We find the presence of \3he as an impurity in hcp solid \4he has a strong effect on the sharp flux reduction at a concentration-dependent characteristic temperature $T_d$. On the other hand, we find that the presence of \3he does not alter the universal temperature dependence of the limiting mass flux above $T_d$.  The magnitude of the flux is typically sample-dependent, and sample-dependent at any concentration, but the temperature dependence is universal.  
 The specific reason for the universal temperature dependence for $T > T_d$ remains unresolved. It is likely due to the physics associated with the conducting pathways.  The results also suggest that the presence of \3he does not destroy the apparent Luttinger-like behavior of the flux.  More experimental and theoretical study of solid helium and 1D superfluidity\cite{Ikegami2007,Toda2007,Kulchytskyy2013} and its pressure dependence are needed.

\section{ACKNOWLEDGMENTS}

We appreciate the many conversations with colleagues in the field, particularly B.V. Svistunov and W.J. Mullin. We also acknowledge the early work with the apparatus by M.W. Ray.  This work was supported by the National Science Foundation grant No. DMR 12-05217  and by Research Trust Funds administeered by the University of Massachusetts Amherst.

\bibliography{ref3a}

\end{document}